\documentclass[]{spie}  %>>> use for US letter paper
\usepackage{amsmath,amsfonts,amssymb}
\usepackage{graphicx}
\usepackage[colorlinks=true, allcolors=blue]{hyperref}

\usepackage{booktabs} % Horizontal rules in tables
\usepackage{comment}
\usepackage[export]{adjustbox}

\title{Contrast performance of an 8m off-axis, segmented space telescope equipped with an adaptive optics system}

\author[a]{Axel Potier}
\author[a]{Garreth Ruane}
\author[b]{Kiarash Tajdaran}
\author[c]{Chris Stark}
\author[a]{Pin Chen}
\author[b]{Larry Dewell}
\author[c,d]{Roser Juanola-Parramon}
\author[b]{Alison Nordt}
\author[e]{Laurent Pueyo}
\author[a]{David Redding}
\author[a]{A J Eldorado Riggs}
\author[f]{Dan Sirbu}

\affil[a]{Jet Propulsion Laboratory, California Institute of Technology, 4800 Oak Grove Drive, Pasadena, CA 91109}
\affil[b]{Lockheed Martin Space, Advanced Technology Center, 3251 Hanover Street, Palo Alto, CA 94304}
\affil[c]{NASA Goddard Space Flight Center, 8800 Greenbelt Rd, Greenbelt, MD 20771}
\affil[d]{Center for Space Sciences and Technology, University of Maryland, Baltimore County, Baltimore, MD 21250}
\affil[e]{Space Telescope Science Institute, 3700 San Martin Dr, Baltimore, MD 21218}
\affil[f]{NASA Ames Research Center, Space Science \& Astrophysics Branch, Moffett Field, Mountain View, CA 94035}

\authorinfo{Send correspondence to Axel Potier: \href{mailto:axel.q.potier@jpl.nasa.gov}{axel.q.potier@jpl.nasa.gov}}%\\ © 2022. All rights reserved. }

\begin{document} 
\maketitle

\begin{abstract}
The Astro2020 decadal survey recommended an infrared, optical, ultra-violet (IR/O/UV) telescope with a $\sim$6~m inscribed diameter and equipped with a coronagraph instrument to directly image exoEarths in the habitable zone of their host star. A telescope of such size may need to be segmented to be folded and then carried by current launch vehicles. However, a segmented primary mirror introduces the potential for additional mid spatial frequency optical wavefront instabilities during the science operations that would degrade the coronagraph performance. A coronagraph instrument with a wavefront sensing and control (WS\&C) system can stabilize the wavefront with a picometer precision at high temporal frequencies ($>$1Hz). In this work, we study a realistic set of aberrations based on a finite element model of a slightly larger (8m circumscribed, 6.7m inscribed diameter) segmented telescope with its payload. We model an adaptive optics (AO) system numerically to compute the post-AO residuals. The residuals then feed an end-to-end model of a vortex coronagraph instrument. We report the long exposure contrast and discuss the overall benefits of the adaptive optics system in the flagship mission success.
\end{abstract}


\section{Introduction}
The 2020 Astronomy and Astrophysics decadal survey recommended a $\sim$6~m infrared, optical, ultra-violet (IR/O/UV) telescope dedicated to the direct imaging of Earth analogue candidates around nearby stars. The LUVOIR-B mission concept is currently the most developed NASA concept of similar scale. LUVOIR-B is an off-axis, segmented telescope with a diameter of 8~m circumscribed and 6.7~m inscribed for a total collecting area of 43.4m$^2$ (55 segments whose size is 0.955m flat-to-flat). The segmented architecture is assumed to be prone to mid- to high-order dynamical aberrations that occur during the science operations that would degrade the performance of following coronagraph instruments used to directly image exoplanets. Adaptive wavefront sensing and control (WS\&C) algorithms are therefore likely to be considered to stabilize the perturbations and provide more favorable observing conditions to the coronagraph.

In this paper, we present simulations of realistic dynamical aberrations and their residuals when using an adaptive optics (AO) system. In Sec.~\ref{sec:statistic}, we describe three different time series of aberrations computed with a finite element model of LUVOIR-B at Lockheed Martin. Section~\ref{sec:performance} presents the post-AO performance of the primary coronagraph architecture proposed for LUVOIR-B.

\section{Statistics of the dynamical wavefront}
\label{sec:statistic}
Following previous work for the LUVOIR-A mission concept \cite{LMC2019}\,, Lockheed Martin Space (LMS) has developed an integrated structural, optical, controls, and noise/disturbance source model to predict the dynamic optical performance of the LUVOIR-B architecture \cite{Tajdaran2022}\,. This integrated modeling tool assumes a vibration isolation and telescope precision pointing system consistent with an LMS-developed Disturbance Free Payload non-contact pointing and vibration isolation architecture, which is designed to achieve a high level of stability in the presence of various disturbance sources (actuator disturbance, sensor noise, interface coupling). The model predicts the optical performance of the observatory, specifically the dynamic WFE (DWFE) by mapping optical node perturbations into an optical path difference (OPD) through a linear optical model (LOM).

\begin{figure}[t]
    \centering
	\includegraphics[width=15cm]{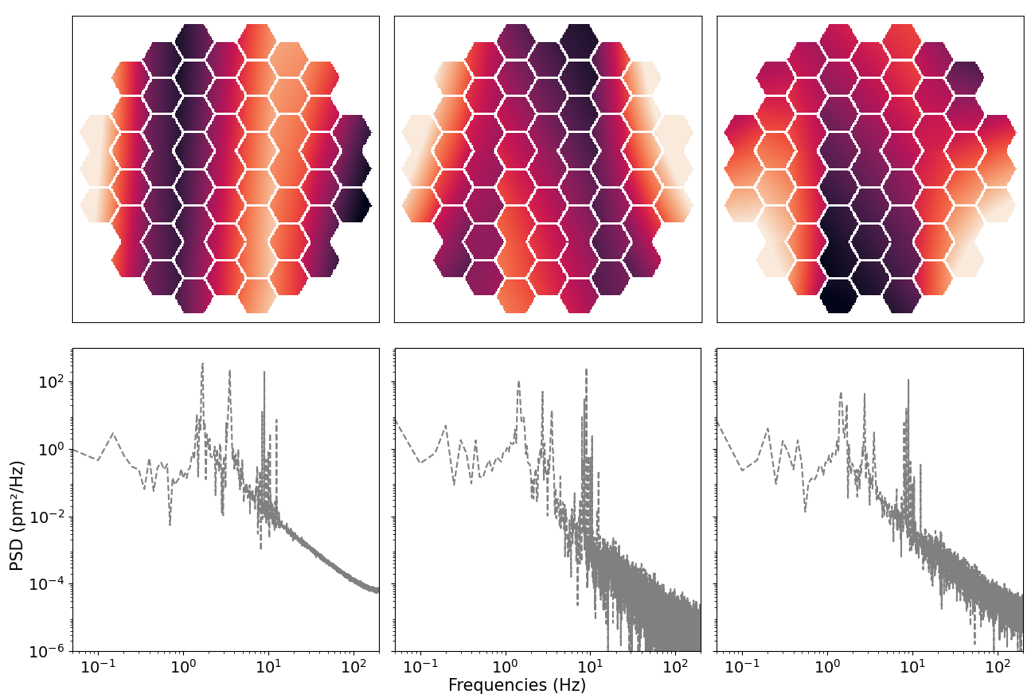}
	\caption{The first 3 principal components of sample 1. (top) The spatial modes and (bottom) their respective PSDs.} % Figure caption
	\label{fig:PCA_PSD} % Label for referencing with \ref{bear}
\end{figure}

The results are 20-second time series sampled with 8000 OPD maps that represent the wavefront errors at the exit pupil of the optical telescope assembly (OTA), or equivalently at the coronagraph instrument entrance pupil. Three different levels of time-varying optical aberrations were generated: 11pm, 20pm and 32pm rms, which we will refer to as sample 1, 2 and 3 respectively. In Fig.~\ref{fig:PCA_PSD}, we show the spatial and temporal distribution of the first three modes of sample 1 that were obtained with a principal component (PC) analysis. Comparing with the decomposition for LUVOIR-A in our previous work\cite{Potier2021}\,, LUVOIR-B's PCs are spatially lower order and the high-order structural modes caused by the unfolding of the primary mirror seen in the LUVOIR-A decomposition are not prominent. Moreover, the three first modes represent 98.5\% of the total variance for LUVOIR-B while the LUVOIR-A integrated model required the combination of 10 modes to explain a similar amount of variance. The vibrating frequencies come primarily from the sunshield and the primary mirror segments.

\section{Post-AO coronagraph performance}
\label{sec:performance}

\begin{figure}[t]
    \centering
    \begin{minipage}[t]{12cm}
    \centering
    \includegraphics[width=\linewidth,valign=t]{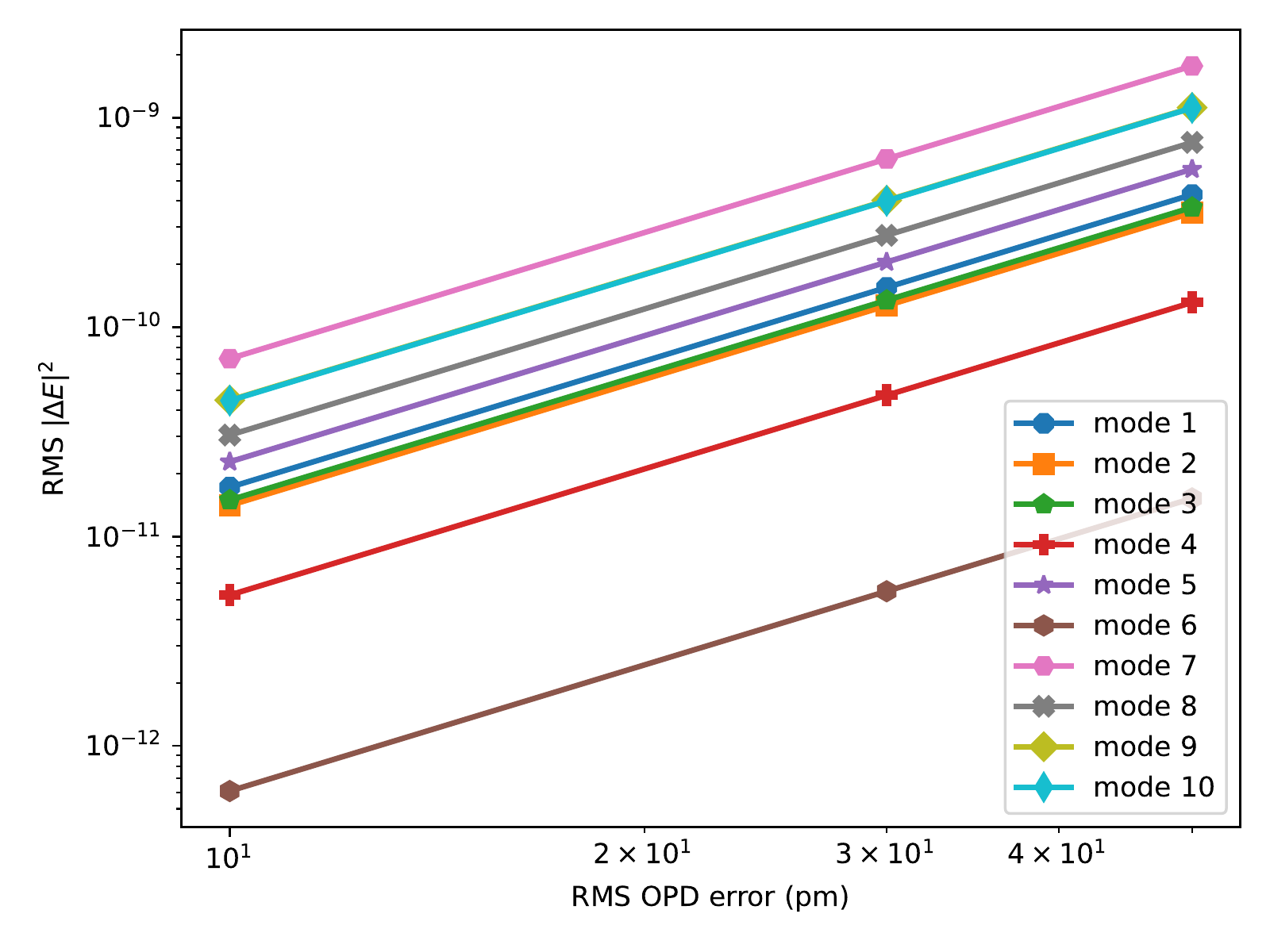}
    \end{minipage}
    \hspace{0.01\textwidth}
    \begin{minipage}[t]{2.85cm}
    \centering
    \includegraphics[width=\linewidth,valign=t]{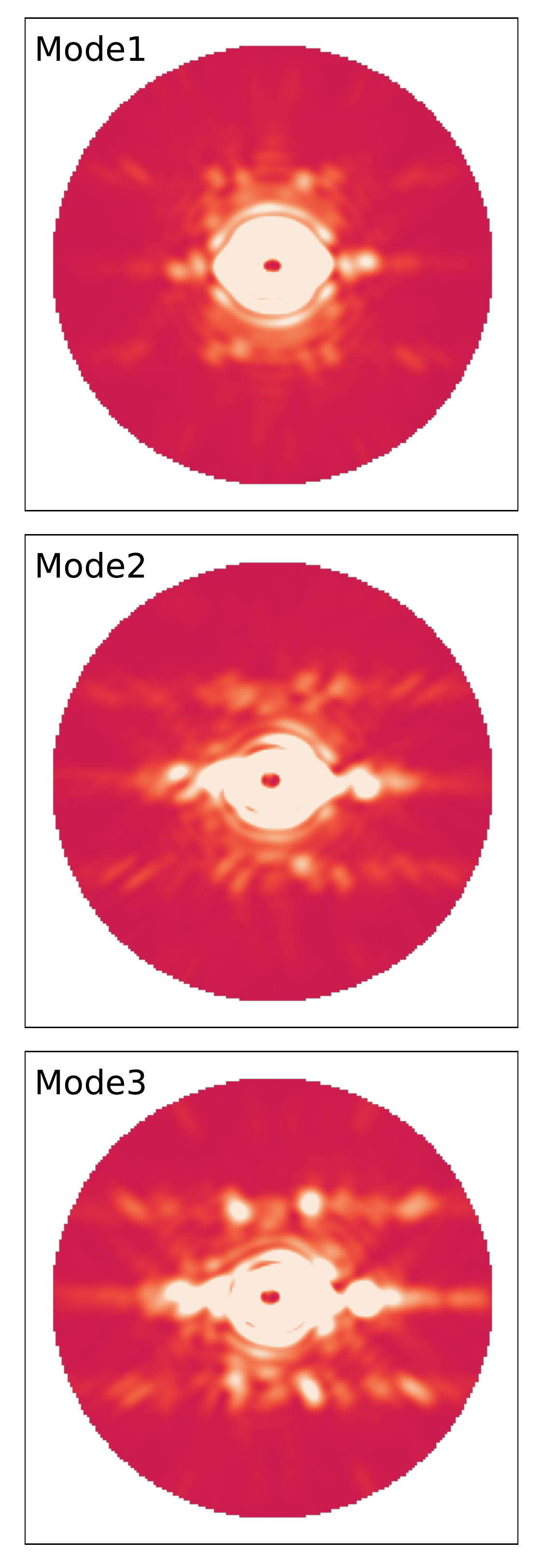}
    \end{minipage}
	\caption{Left: The root-mean-square (RMS) of the square magnitude of the change in electric field in the dark zone, $\Delta E$, with respect to the RMS of the sample 1's PCA modes introduced in the vortex charge 6 coronagraph entrance pupil. Right: spatial distribution of the intensity induced by the first three modes in the DH. The ideal diffraction pattern of the vortex coronagraph (i.e. with no aberration) has been subtracted.} % Figure caption
	\label{fig:modal_images} % Label for referencing with \ref{bear}
\end{figure}

Using the methods developed in our previous work for LUVOIR-A \cite{Potier2021}\,, we assess the post-AO coronagraph performance of LUVOIR-B. We first analytically model the temporal response of a simple AO system that includes a wavefront sensor (WFS), a controller (either integrator or predictive), and a digital to analogue converter. The sampling frequency of the WFS is either 100~Hz or 1000~Hz and the time delay due to computing calculation is set to one frame (i.e. 10ms or 1ms respectively). We assume a Zernike wavefront sensor (ZWFS) since it provides a theoretically high sensitivity in the photon noise limit \cite{Guyon2005}\,. Non-common-path aberrations between the science and sensing path are considered to be nulled or perfectly corrected by the deformable mirror. The spectral bandwidth for sensing is $\Delta\lambda/\lambda$=20\% in the visible, the throughput of the entire ZWFS path is assumed to be 30\%, and the telescope is a circular and unobscured with a collecting area whose diameter is 7~m. We assume no cross-talk between the modes sensed by the ZWFS (i.e. that the sampling is sufficient to differentiate between the spatial modes). The modal gains are then optimized in parallel for the first ten PCs to minimize their individual post-AO variance at each guide star magnitude.

The post-AO statistics of the DWFE is then feed to a end-to-end model of coronagraph. We here adopt a deformable mirror-assisted apodization vortex coronagraph (DM-AVC) with a topological charge of 6 \cite{Ruane2018d,Pueyo2019}\,. Indeed, vortex coronagraphs have smaller inner working angles (IWA) and higher throughput than apodized pupil lyot coronagraphs (APLC) solutions. In the absence of central obscuration, they are naturally robust to stellar angular size and low order aberrations up to spherical while a charge 8 or more could even be considered to keep improving the insensitivity to low orders. The LUVOIR-B off-axis architecture therefore favors vortex coronagraphs. The combination of such coronagraphs with DM assisted apodization enables the correction of diffraction caused by the gaps between mirror segments ($<0.1$\% of the pupil diameter)\cite{Ruane2018d}\,. The DM shape is optimized through a FALCO compact model\cite{Riggs2018} to minimize the starlight leakage in an annulus whose angular separations ranges from $3\lambda/D$ to $28\lambda/D$ (a.k.a. the dark zone) at seven discrete wavelengths over a 10\% spectral bandwidth.

Figure~\ref{fig:modal_images} shows the impact of each structural mode on the dark zone in the science image. Modes 1, 2 and 3 demonstrate a similar contribution to the error budget for an equal level of perturbation. A time series that would be dominated by modes 4 and 6 (that respectively look like astigmatism and spherical Zernike modes) would provide even more favorable conditions for the vortex coronagraph. On the other hand, other modes that include higher spatial frequencies like mode 7 would have a larger impact on the instrument performance, as expected. Each mode has a proportional relationship between the intensity leakage on the science detector and the variance of each individual input modes. We can therefore compute the normalized intensity map as:
\begin{equation}
\label{eq:MeanContrast_calculation}
    I = |E_0|^2\ + \sum_{i=1}^{n}\langle a_i^2 \rangle|\Delta E_i|^2, 
\end{equation}
where $|E_0|^2$ stands for the coronagraph's unaberrated point spread function while $\Delta E_i$ and $\langle a_i^2 \rangle$ represent the induced stellar leakage on the science detector and the mean variance of the structural mode $i$ over the post-AO time series, respectively. This simplification expedites the performance calculation by propagating 10 modes to determine $\Delta E_i$ instead of propagating 8000 OPD maps through the coronagraph end-to-end model.

\begin{figure}[t]
    \centering
	\includegraphics[width=12cm]{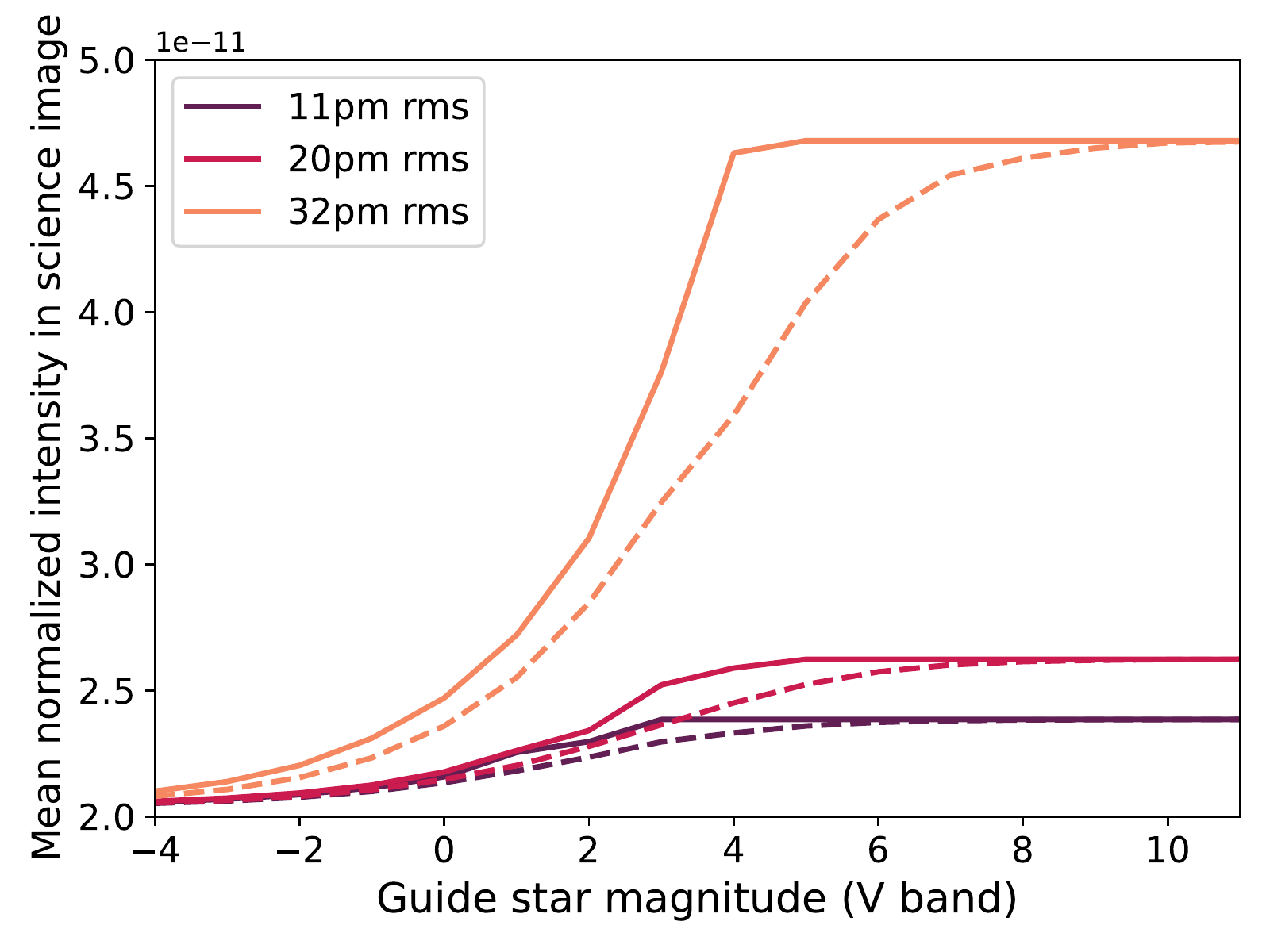}
	\caption{Optimized post-AO normalized intensity in the dark zone with respect to the magnitude of the guide star, for the three different LUVOIR-B input optical aberration data sets. Dashed lines represent predictive control while continuous lines represent the integrator case.} % Figure caption
	\label{fig:comparison} % Label for referencing with \ref{bear}
\end{figure}
Figure~\ref{fig:comparison} shows the post-AO coronagraph performance in the dark zone of the science image with respect to the guide star magnitude and for each time series. First, the dynamical aberrations remain uncorrected in any case above a magnitude of 8, but the uncorrected performance is well below the 1e-10 contrast that is usually targeted by coronagraph designers. For the worst case scenario that was simulated (32pm rms of input aberrations), the normalized intensity is only degraded by a factor of 2.3. 

Assuming the linear relationship between the normalized intensity in the dark zone and the modal variance as demonstrated in previous work \cite{Potier2021}\,, we should be able to increase the dynamical aberration error budget to $\sim$55pm rms without AO correction while preserving a normalized intensity under $10^{-10}$. This will be investigated with newly generated time series in future work. If viable, the OPD requiremens would be relaxed by a factor 5 in standard deviation (25 in variance) with respect to what was found for the combination of LUVOIR-A and the APLC\cite{Potier2021}\,. The improved performance is due to the LUVOIR-B structural modes being dominated by spatially low-order components that are well rejected by the DM-AVC.

The initial contrast achieved with the three time series considered above would make the AO correction unnecessary. Indeed, former yield calculations demonstrate a weak dependence on the raw normalized intensity if a low ($<$26.5~mag) noise floor (faintest detectable source) is attained \cite{Stark2019}\,. From our simulations, even the worse case scenario should provide a noise floor under 26.5~mag (3e-11 of normalized intensity) after post-processing since only a factor of 2 of correction with respect to the raw normalized intensity would be required. In any circumstance, the obtained yield is therefore close to the ideal yield for LUVOIR-B, formerly estimated at $\sim25$ exoEarth candidates, assuming 2 years of telescope time\cite{Stark2019}\,. However, AO could be used to further relax the contribution of dynamical aberrations in the error budget. 

Comparing sample 1 (11~pm~RMS) for LUVOIR-B and sample B ($\sim$10~pm~RMS) for LUVOIR-A\cite{Potier2021}\,, an integrator can be used with fainter stars in the case of LUVOIR-B (mag$<$2 vs mag$<$0 for LUVOIR-A for the $\sim10$~pm RMS case) despite collecting 4 times less photons with the WFS. This is because of the higher signal to noise ratio as well as the frequencies of vibrations being lower than 10~Hz for the first three modes, which provides more favorable conditions to the integrator. Conversely, the multiplicity of vibrations at equal amplitude penalizes predictive control to mitigate one or two vibrations with a low amount of photons because of the Bode's sensitivity integral theorem. Therefore, a correction can be applied for mag$<$6 natural guide star compared with mag$<$8 for the case of LUVOIR-A.

\section{Conclusion}
Building on our previous work for LUVOIR-A\cite{Potier2021}, we performed end-to-end analytical modeling of the contrast performance provided by an adaptive optics system working under LUVOIR-B realistic observing scenarios and turbulent regimes. We showed that under the worse-case conditions (32pm rms at the exit pupil of the OTA) and up to a derived perturbation of 55pm rms, the contrast remains well under the "targeted" normalized intensity of $10^{-10}$ assumed to be required to detect a large sample of exoEarth candidates. Comparing those results with LUVOIR-A, we conclude that the input variance of the aberrations alone is insufficient to derive yield performance. Indeed, $\sim$10~pm RMS of aberrations provides a contrast 4 times better for LUVOIR-A than for LUVOIR-B. Instead, the spatial distribution of the architecture structural modes needs to be taken into account in combination with the coronagraph choice. Then, two solutions can be considered to preserve the ideal yield during the mission. Either the coronagraph is designed to adapt its null space to the input aberrations or the telescope architecture is optimized according to the null space of already existing coronagraphs. In both cases, the coronagraph and telescope should be designed and treated as a single system.

\acknowledgments % equivalent to \section*{ACKNOWLEDGMENTS}       
The research was carried out at the Jet Propulsion Laboratory, California Institute of Technology, under a contract with the National Aeronautics and Space Administration (80NM0018D0004). 

% References
\bibliography{bib_GS}   %>>>> bibliography data in report.bib
\bibliographystyle{spiebib} % makes bibtex use spiebib.bst

\end{document}